\documentclass[twocolumn, final]{IEEEtran}
\usepackage{amsfonts}
\usepackage{amsmath,cases,amssymb}
\usepackage[dvips]{graphicx}
\usepackage{cite,color}
\usepackage{color}
\usepackage{subcaption}
\usepackage{epsfig}
\usepackage{epstopdf}
\usepackage{url}
\usepackage{algorithm}
\usepackage{algorithmicx, algpseudocode}
\usepackage{color}
\usepackage{multirow}
\usepackage{mathtools}
\usepackage{balance}

\newcommand{\ds}{\displaystyle}

\newcommand{\clA}{{\cal A}}
\newcommand{\clD}{{\cal D}}

\newcommand{\clB}{{\cal B}}

\newcommand{\clH}{{\cal H}}
\newcommand{\clN}{{\cal N}}

\newcommand{\qed}{$\quad \Box$}
\newcommand{\Tr}{{\rm Trace}}

\newtheorem{myth}{Theorem}
\newtheorem{mypro}{Proposition}
\newtheorem{mylem}{Lemma}

\newcommand{\bgeqn}{\begin{equation}}
\newcommand{\edeqn}{\end{equation}}
\newcommand{\la}{\langle}
\newcommand{\ra}{\rangle}
\newcommand{\lan}{\langle}
\newcommand{\ran}{\rangle}

\newcommand{\beqa}{\begin{eqnarray}}
\newcommand{\eeqa}{\end{eqnarray}}
\newcommand{\beqas}{\begin{eqnarray*}}
\newcommand{\eeqas}{\end{eqnarray*}}

\newcommand{\bX}{\mathbf{X}}

\newcommand{\clR}{{\cal R}}
\newcommand{\bx}{\pmb{x}}

\newcommand{\clI}{{\cal I}}

\newcommand{\ba}{\mathbf{a}}
\newcommand{\bb}{\mathbf{b}}

\newcommand{\Prf}{{\it Proof. }}
\newcommand{\bT}{\mathbf{T}}
\newcommand{\bY}{\mathbf{Y}}
\newcommand{\polyD}{{\sf Poly}(\clD_S)}
\newcommand{\inte}{{\sf int}}
\begin{document}
\title{PMU Placement Optimization for Smart Grid Obvervability and State Estimation\thanks{This work was supported in part by the U.S. National Science Foundation under Grants CNS-1702808 and DMS-1736417.}}
\author{Y. Shi, H. D. Tuan, A. A. Nasir, T. Q. Duong, and H. V.  Poor
\thanks{Ye Shi and Hoang D. Tuan are with the School of Electrical and Data Engineering, University of Technology Sydney, Broadway, NSW 2007, Australia (email: Ye.Shi@student.uts.edu.au, Tuan.Hoang@uts.edu.au)}
\thanks{Ali A. Nasir is with the Department of Electrical Engineering, King Fahd University of Petroleum and Minerals (KFUPM), Dhahran, Saudi Arabia (email: anasir@kfupm.edu.sa).}
\thanks{Trung Q. Duong is with Queen's University Belfast, Belfast BT7 1NN, UK  (email: trung.q.duong@qub.ac.uk)}
\thanks{H. Vincent Poor is with the Department of Electrical Engineering, Princeton University, Princeton, NJ 08544, USA (email: poor@princeton.edu)}
}
\date{}
\maketitle
\begin{abstract} In this paper, phasor measurement unit (PMU) placement for power grid  state estimation under
different degrees of observability is studied. Observability degree  is the depth of
the buses' reachability by the placed
PMUs and thus constitutes an important characteristic for PMU placement.
However, the sole observability as addressed in many works still does not guarantee a good estimate for the grid state. Some existing works also considered the PMU placement for minimizing
the mean squared error or maximizing the mutual information between the measurement output and grid state. However, they ignore
the obsvervability requirements for computational tractibility and thus potentially lead to artificial results such as acceptance of the estimate for
an unobserved state component as its unconditional mean. In this work, the PMU placement optimization problem is considered by
minimizing the mean squared error or maximizing the mutual information between the measurement output and grid state, under grid observability constraints. The provided solution is free from the mentioned fundamental drawbacks in the existing PMU placement designs. The problems are posed as  binary nonlinear optimization problems, for which this paper develops efficient algorithms for computational solutions. The performance of the proposed algorithms is analyzed in detail through numerical examples on large scale IEEE power networks.
\end{abstract}
%\vspace{-1cm}
\section{Introduction}
Phasor measurement unit (PMU) is an advanced digital meter, which is used in smart power grids for real-time monitoring of grid operations \cite{DCTP10}. By installing it at a buse, the state-of-the-art PMU can  measure not only the phasor of the bus voltage but also the current phasors of incident power branches with high accuracy \cite{PT08}. These measurements are explored by the modern energy management systems (EMSs) for critical applications such as  optimal power flow, contingency analysis, and cyber security, etc. \cite{MAE99,ZZDK16,TGB16}.

As phasor measurement units (PMUs) are costly, there is a vast amount of literature on PMU placement optimization
to target the minimal number of PMUs. Under different degrees of observability, the mission is accomplished
by binary linear programming (BLP) \cite{G108,G208}. Here, the complete observability means that there is no
bus left unobserved by the placed PMUs, while depth-of-$n$ unobservability means that there are
at most $n$ connecting buses left unobserved by the placed PMUs \cite{NP05}, making as many states as possible
observed by restricted number of PMUs. An exhaustive binary search was proposed in \cite{CK08} to deal with this objective
 under the complete observability condition and
additional operating conditions such as the single branch outage and the presence of  zero power injections.
A binary particle swarm optimization algorithm was proposed in
\cite{HRAA11}  to deal with it while maintaining  the complete observability conditions under the contingencies of PMU loss or branch outage. Binary quadratic programming and BLP were respectively used in \cite{CKE09} and \cite{KBRG17} to
study the effect of conventional measurements and zero bus injections to the complete observability.

Apparently, observability alone does not necessarily lead to a meaningful state estimate or an informative PMU configuration. In fact, PMU configurations, which use the same number of PMUs to make the grid completely observable, can result in quite different estimation accuracies \cite{RH06}. Intuitively, a better estimator can be obtained by
appropriately employing more PMUs.
PMU placement optimization to minimize the mean squared error (of grid state estimation) or to maximize
the mutual information between the measurement output and grid state under a fixed allowable number of PMUs was considered
in \cite{KGW12} and \cite{Lietal13}, respectively. Obviously, these placement tasks are mathematically modelled by
optimization of nonlinear objective functions of binary variables
subject to a simple linear constraint for fixing the number of PMUs.
A convex relaxation with the binary constraint $\{0,1\}$ for
binary variables relaxed to the box constraint $[0,1]$ was proposed in \cite{KGW12}, which not only fails to provide even a local
optimal solution in general but also is not scalable in the grid dimension as it involves an additional
large-size semi-definite matrix variable. A greedy algorithm proposed in \cite{Lietal13} does not provide a local optimal
solution  either. More importantly, both \cite{KGW12} and \cite{Lietal13} ignored
observability constraints for computational tractability. It was argued in \cite{Lietal13} that its proposed mutual information criterion includes the grid complete observability, which is obviously not right simply because as
shown later in the paper,  the
latter differentiates the state estimate from its unconditional mean, which is the trivial estimate,
while the former does not.

To fill the gap due to disconnected considerations for the grid state observability and state estimation
in the existing approaches, this paper considers PMU placement to optimize the estimation performance under different degrees
of observability and with a fixed number of PMUs.   These problems are posed as  binary nonlinear optimization problems,
which are computationally much challenging. To the authors' best knowledge,
such optimization problems are still quite open for research.

The rest of the paper is structured as follows. Section
II is devoted to the problem statement, which also particularly shows the importance of imposing observability constraints
in optimization formulations. Section III develops two scalable algorithms for PMU placement
optimization to minimize the mean squared error (of grid state estimation)
or maximize the mutual information between the measurement outputs and phasor states under a fixed number of PMUs
and different degrees of observability. Section IV presents tailored path-following discrete optimization algorithms for the problems without observability constraint. Simulations are provided in Section V, which demonstrates the efficiency of our algorithms. Section VI concludes the paper. The fundamental inequalities used in Section III are given in the Appendix.\\

{\it Notation.} The notation used in this paper is standard. Particularly, $A\succ 0$ ($A\succeq 0$, resp.) for a Hermitian symmetric matrix $A$ means that it is positive definite (semi-definite, resp.).
$\Tr(.)$ and $|.|$ are the trace and determinant operator.
$1_N$ is an $N$-dimensional vector of ones. $I_N$ is the identity matrix of size $N$.
$a\leq b$ for two real vectors $a=(a_1,\dots, a_n)^T$ and $b=(b_1, \dots, b_n)^T$ is componentwise understood, i.e.
$a_i\leq b_i$, $i=1, \dots, n$. The cardinality of a set ${\cal C}$ is denoted by $|{\cal C}|$.
$\mathbb{E}(.)$ denotes expectation, so the mean $\bar{u}$ of a random variable (RV) $u$ is $\bar{u}=\mathbb{E}(u)$.
For two random variables $u$ and $v$, their cross-covariance matrix $R_{uv}$ is $\mathbb{E}((u-\bar{u})(v-\bar{v})^T)$.
Accordingly,  the autocovariance $\clR_u$ of $u$ is $\mathbb{E}((u-\bar{u})(u-\bar{u})^T)$.   $u\sim\clN(\bar{u}, \clR_{u})$ means $u$ is a Gaussian random variable with means $\bar{u}$ and autocovariance $\clR_{u}$, which represent the first moment of $u$. The entropy of $u$ is
 $\clH(u)=\frac{1}{2}\log_2|\clR_u|=\frac{1}{2\ln 2}\ln|\clR_u|$. Finally, denote by $u|v$  a RV $u$ conditioned on the RV $v$.
\section{Problem statement}
Consider a power grid with a set of buses indexed by ${\cal N}:=\{1,2,\cdots,N\}$, where buses are connected through a set of transmission lines ${\cal L}\subseteq {\cal N}\times {\cal N}$, i.e. bus $k$ is connected to bus $m$ if and only if $(k,m)\in {\cal L}$. Accordingly, $\clN(k)$ is the set of other buses connected to bus $k$.
In a DC power model, the power injection at bus $k$ is approximated by
\begin{eqnarray}\label{p_inj}
P_k = B_{kk}\theta_k+\sum_{m\in \clN(k)}B_{km}\theta_m,
\end{eqnarray}
where $P_k$ is the power injection at bus $k$ and $\theta_m$ is the voltage phasor angle at bus $m$, while
$B_{km}$ is the imaginary part of the $(k,m)$-entry of the grid's admitance matrix $Y$.
Let $P:=(P_1,\dots,P_N)^T\in\mathbb{R}^N$ be the power injection vector and
$\theta:=(\theta_1, \dots, \theta_N)^T\in\mathbb{R}^N$ be the voltage phasor vector. Then (\ref{p_inj}) can be re-written as
$P= B \theta$, where $B\in\mathbb{R}^{N\times N}$ is the so called susceptance matrix with the entries
$B(k,k)=B_{kk}$ and $B(k,m)=B_{km}$, if $m\in\clN(k)$, while
$B(k,m)=0$, otherwise. The susceptance matrix $B$ is invertible under the assumption that the grid is fully connected \cite{GKP80}. Since
$P$ can be assumed to be $\clN(u_p,\Sigma_P)$ \cite{AWJ05}, it is obvious that
$\theta\sim \clN(B^{-1}u_p, B^{-1}\Sigma_p(B^{-1})^T)$.

On the other hand, the measurement equation of a PMU installed at bus $k$ in the linear DC power flow model \cite{GJ1994} is \cite{PT08,Reetal10,Lietal13},
\begin{equation}\label{se1}
\begin{array}{lll}
  \zeta_k&=& \theta_k + \vartheta_k,  \\
  \zeta_{km} &= & \theta_k - \theta_m + \vartheta_{km},\quad k\in \clN, m\in\clN(k),
\end{array}
\end{equation}
with noises $\vartheta_k\sim \clN(0,r_k)$ and $\vartheta_{km}\sim\clN(0,\rho_k)$.
The number of incident lines of bus $k$ is the cardinality $|\clN(k)|$. Accordingly, the measurement vector
$z_k:=(\zeta_k, \zeta_{k1}, \dots, \zeta_{k|\clN(k)|})^T$ is of dimension $M_k = |\clN(k)|+1$. For simplicity, (\ref{se1}) is rewritten in regression form as:
\begin{eqnarray}\label{se2}
z_k = H_k \theta+w_k,
\end{eqnarray}
where $H_k \in \mathbb{R}^{M_k\times N}$ is the associated regression matrix,
$w_k:=(\vartheta_k, \vartheta_{k1}, \dots, \vartheta_{k|\clN(k)|})^T \sim \clN(0,R_{w_k})$ with diagonal covariance $R_{w_k}$.

To describe the presence or absence of PMU at bus $k$, we introduce a selection vector
$\bx=(x_1,\cdots,x_N)^T \in\{0,1\}^N$,
where $x_k = 1$ if a PMU is installed at bus $k$, and  $x_k = 0$ otherwise.
Let us assume that we have $S$ PMUs in total for installation, so
\begin{eqnarray}\label{conbx}
\sum_{k\in\clN}x_k = S.
\end{eqnarray}
Define
\begin{equation}\label{sens_num}
\clD_S:=\{ \bx\in \{0,1\}^N\ :\ \sum_{k\in\clN}x_k=S\}
\end{equation}
and $\bX = \mbox{diag}[x_k\clI_k]_{k=1, \dots,N}$, $\clR_{w}=\mbox{diag}[R_{w_k}]_{k\in\clN}$,
 where $\clI_k$ is the identity matrix of size $M_k\times M_k$.

For every $\bx\in\clD_S$, let $k_j\in\clN$, $j=1, \dots, S$ for which $x_{k_j}=1$. Define accordingly,
$\clR_{w}(\bx)=\mbox{diag}[\clR_{w_{k_j}}]_{j=1, \dots, S}$, and
\[
\begin{array}{c}
z(\bx)=\begin{bmatrix}
z_{k_1}\\
\cdots\\
z_{k_S}\end{bmatrix}, \
w(\bx)=\begin{bmatrix}w_{k_1}\cr
\cdots\cr
w_{k_S}\end{bmatrix},\
\bar{H}(\bx)=\begin{bmatrix}H_{k_1}\\
\cdots\\
H_{k_S}\end{bmatrix}.
\end{array}
\]
The multi-input-multi-output PMU measurement equation is
\[
z(\bx) = \bar{H}(\bx)\theta+w(\bx).
\]
It is obvious that $\clR_{z(\bx)\theta} = \bar{H}(\bx)\clR_{\theta}$ while
$\clR_{z(\bx)} = \bar{H}(\bx)R_{\theta}\bar{H}(\bx)^T +R_{w(\bx)}$. Let $\theta|z(\bx)$ be the RV
$\theta$ conditioned on the RV $z(\bx)$. By \cite{P94}
\begin{equation}\label{condi}
\theta|z(\bx)\sim\clN(\hat{\theta},\clR_e(\bx)),
\end{equation}
where
\[
\begin{array}{lll}
\hat{\theta} &=& \bar{\theta} + \clR_{z(\bx)\theta}^T\clR_{z(\bx)}^{-1}(z(\bx)-\overline{z(\bx)})\nonumber\\
&=& \bar{\theta} + \clR_{\theta}\bar{H}(\bx)^T(\bar{H}(\bx)\clR_{\theta}\bar{H}(\bx)^T + \clR_{w(\bx)})^{-1}\nonumber\\
&&\times(z(\bx) - \bar{H}(\bx)\bar{\theta}),
\end{array}
\]
which is the minimum mean squared error (MMSE) estimate of $\theta$ based on PMU output
$z(\bx)$, and
\begin{eqnarray}
\clR_e(\bx) &=& \clR_{\theta} - \clR_{z(\bx)\theta}^T \clR_{z(\bx)}^{-1}\clR_{z(\bx)\theta}\nonumber\\
&=& \ds\clR_{\theta} - \clR_{\theta}\bar{H}(\bx)^T\left(\bar{H}(\bx)\clR_{\theta}\bar{H}(\bx)^T +\clR_{w(\bx)}\right)^{-1}\nonumber\\
&&\times \bar{H}(\bx)\clR_{\theta}\nonumber\\
&=&\left(\clR_{\theta}^{-1} + \bar{H}(\bx)^T \clR_{w(\bx)}^{-1}\bar{H}(\bx)^T\right)^{-1}\nonumber\\
&=&\ds \left(\clR_{\theta}^{-1} +\sum_{j=1}^{S}H_{k_j}^T\clR_{w_{k_j}}^{-1}H_{k_j}\right)^{-1}\\
&=& \left(B^T\Sigma_P^{-1}B + \ds\sum_{k\in\clN}x_k H_k^T\clR_{w_k}^{-1}H_k\right)^{-1}.\label{clRe}
\end{eqnarray}
The mean squared error (MSE) $\mathbb{E}(||\theta-\hat{\theta}||^2)$ is
\[
f_{e}(\bx):=\Tr(\clR_e(\bx)),
\]
which obviously is an analytical function of the PMU selection vector $\bx$.

Further,  the mutual information (MI) $I(\theta;z(\bx))$ between RVs $\theta$ and $z(\bx)$ is \cite[formula (6)]{Tuetal07}
\[
\begin{array}{lll}
I(\theta;z(\bx))&=&\clH(\theta)-\clH(\theta|z(\bx))\\
&=&\ds\frac{1}{2\ln 2} ( \ln|\clR_{\theta}|-\ln|\clR_e(\bx)|).
\end{array}
\]
Maximizing the MI $I(\theta;z(\bx))$ is thus equivalent to maximizing $f_{MI}(\bx)$ for
\[
f_{MI}(\bx):=-\ln|\clR_e(\bx)|=
\ln|B^T\Sigma_P^{-1}B + \ds\sum_{k\in\clN}x_k H_k^TR_{w_k}^{-1}H_k|.
\]
It should be realized that either the MSE $f_e(\bx)$ or MI $f_{MI}(\bx)$ does not indicate the depth
of the placed PMUs in reaching the measurement for the whole phasor state.  One needs either the
constraint
\begin{equation}\label{obser}
\clA\bx\geq 1_N,
\end{equation}
of the complete observability to assure that the phasor state $\theta$ is completely observable
\cite{Reetal10,KS11,GA13}, where
$\clA$ is the bus-to-bus incidence matrix defined by
$\clA_{km}=1$ if $k = m$ or bus $k$ is adjacent to bus $m$, and $\clA_{km}=0$
otherwise, or the constraint
\begin{equation}\label{depthone}
\clB\clA\bx\geq 1_{N_B},
\end{equation}
of the depth-of-one unobservability to assure that there are no two connecting buses that are unobservable \cite{NP05}. Here and after $\clB$ is the branch-to-bus incident matrix and
$N_B$ is the total number of branches. The general case of dept-of-$n$ unobservability with an arbitrary $n$ is treated similarly though
its practicability is unknown. \\
Let us analyse the constraints (\ref{obser}) and (\ref{depthone}) from the information-theoretic view point. The constraint
(\ref{obser}) guarantees that all state components $\theta_m$ are observable, i.e. each $\theta_m$
appears at least once in the measurement
equations (\ref{se1}), which implies $\theta_m|z(\bx) \neq \theta_m$,
making the measurement equations (\ref{se1}) meaningful for
estimating $\theta_m$. When some $\theta_m$ is not observable, i.e. it does not appear
in the measurement equations (\ref{se1}),
it follows that $\theta_m|z(\bx)=\theta_m$ so the measurement equations in (\ref{se1}) are useless
for estimating $\theta_m$. In this case, the
estimate for $\theta_m$ is its unconditional mean $\bar{\theta}_m$ with
$\mathbb{E}((\theta_m-\bar{\theta}_m)^2)=\clR_{\theta}(m,m)$ and $I(\theta_m;z(\bx))=\clH(\theta)-\clH(\theta|z(\bx))=0$. In other words,
the optimization problem for maximizing $I(\theta;z(\bx))$
does not reveal a nontrivial estimate for $\theta_m$ that is a contradiction to
\cite[statement 1), page 448, 2nd column]{Lietal13} which states that the mutual information metric includes  the complete
observability condition (\ref{obser}) as a special case. Of course, the number of PMUs, $S$, needs to be sufficient enough to make the constraint (\ref{obser}) fulfilled.  When $S$ is not allowed to be
sufficient, one may go for more relaxed constraint (\ref{depthone}), which forces all neighboring buses of any unobservable bus to be observable and thus essentially makes as many states as possible
be observable by the PMUs.

Thus, we can state the problem of PMU placement optimization to minimize the MMSE or to maximize the
MI between the measurement output and phasor state under a fixed number of PMUs and observability/depth-of-one unobservability as the following binary nonlinear optimization problem
\begin{equation}\label{OPP1}
\ds\min_{\bx} f(\bx)\quad
\mbox{s.t.}\quad  \bx\in\clD_S, (\ref{obser})/(\ref{depthone}),
\end{equation}
where $f(\bx)\in\{f_{e}(\bx),-f_{MI}(\bx)\}$, which is a convex function.
\section{Scalable Penalty algorithms for optimal PMU selection}
It is obvious that the main issue is regarding how to handle the discrete constraint $\bx\in\clD_S$ in (\ref{OPP1}). The following
result establishes the equivalence of this discrete constraint  and a continuous constraint.

\begin{mylem}\label{bilem} {\it For a polytope
$\polyD=\{\bx\in [0,1]^N:\ \sum_{k\in\clN}x_k=S\}$,
the discrete constraint $\bx\in\clD_S$ in (\ref{OPP1}) is equivalent to the continuous constraint
\begin{equation}\label{ai}
\bx\in\polyD, g(\bx)\geq S,
\end{equation}
for  $g(\bx):=\sum_{k\in\clN}x_k^L$ with $L>1$.}\\
\end{mylem}
\Prf \ Note that $x_k^L\leq x_k$ $\forall\ x_k\in [0,1]$, so $g(\bx)\leq \sum_{k\in\clN}x_k=S$ $\forall \bx\in\polyD$.
Therefore constraint (\ref{ai}) forces  $g(\bx)=S$, which is
 possible if and only if  $x_k^L=x_k$, $k\in\clN$, i.e $x_k\in\{0,1\}$, $k\in\clN$, implying $\bx\in\clD_S$.\qed\\

Since $g(\bx)$ is convex in $\bx$, the constraint $g(\bx)\geq S$ in (\ref{ai}) is a reverse convex constraint \cite{Tuybook}. As such $\clD_S=\polyD\setminus\{\bx\ :\ g(\bx)<S\}$, i.e.
$\clD_S$ is difference of two convex sets $\polyD$ and $\{\bx\ :\ g(\bx)<S\}$.
Also as $L$ decreases, $g(\bx)$ tends to approach a linear function
$\sum_{k\in\clN}x_k$ and thus,
the constraint $g(\bx)\geq S$ approaches the linear constraint
$\sum_{k\in\clN}x_k\geq S$. However, it does not mean that choosing $L$ closer
to $1$ is effective because the function $g(\bx)-S$ also approaches zero very quickly, making
the constraint $g(\bx)\geq S$ highly artificial. In our previous works
\cite{CTN14,Taetal17}, $L=2$ was chosen.
However, as we will see shortly, $L=1.5$ is a much better choice,  accelerating the convergence of the iterative computational processes. The following result is a direct consequence of Lemma \ref{bilem}.\\

\begin{mypro}\label{bilem1} {\it The function
\[
\tilde{g}(\bx)=1/g(\bx)-1/S
\]
can be used to measure the degree of satisfaction of
the discrete constraint $\bx\in\clD_S$ in the sense that $\tilde{g}(\bx)\geq 0\ \forall\ \bx\in\polyD$
and $\tilde{g}(\bx)=0$ if and only if $\bx\in\clD_S$.} \qed \\
\end{mypro}

Following our previous developments in \cite{CTN14} and \cite{Taetal17}, instead of handling  constraint (\ref{ai}), we incorporate the degree of its satisfaction  into the objective in (\ref{OPP1}), leading to the following penalized optimization problem:
\begin{eqnarray}\label{OPPA3}
\ds\min_{\bx} F_{\mu}(\bx):=f(\bx)+\mu(1/g(\bx)-1/S)\nonumber\\
\mbox{s.t.} \quad \bx\in \polyD, (\ref{obser})/(\ref{depthone}),
\end{eqnarray}
where $\mu>0$ is a penalty parameter. This penalized optimization problem is exact with a sufficiently large $\mu$.
Note that (\ref{OPPA3}) is a minimization of a nonconvex function over a convex set. We now develop a path-following
computational procedure for its solution. For this purpose, we firstly develop an upper bounding approximation for (\ref{OPPA3}), at some feasible point $\bx^{(\kappa)}$ (at $\kappa$-th iteration).
As the function $g(\bx)$ is convex, it is true that \cite{Tuybook},
\[
\begin{array}{lll}
g(\bx)&\geq&g^{(\kappa)}(\bx)\\
&:=& g(\bx^{(\kappa)})+\la \nabla g(\bx^{(\kappa)}),\bx-
\bx^{(\kappa)}\ra\\
&=&-(L-1)\ds\sum_{k\in\clN}(\bx_k^{(\kappa)})^L+
L\ds\sum_{k\in\clN}(\bx_k^{(\kappa)})^{L-1}\bx_k.
\end{array}
\]
Therefore, an upper bounding approximation at $\bx^{(\kappa)}$ for $1/g(\bx)$ can be easily obtained as
$1/g(\bx)\leq 1/g^{(\kappa)}(\bx)$ over the trust region
\begin{equation}\label{quad2b}
g^{(\kappa)}(\bx)>0.
\end{equation}
At the $\kappa$-th iteration we are supposed to solve the following convex optimization problem to generate the
next iterative point $\bx^{(\kappa+1)}$:
\begin{eqnarray}\label{OPPA4}
\ds\min_{\bx} f(\bx)+\mu(1/g^{(\kappa)}(\bx)-1/S)\nonumber\\
\mbox{s.t.}\ \bx\in\polyD, (\ref{obser})/(\ref{depthone}), (\ref{quad2b}).
\end{eqnarray}
Although function $f(\bx)$ is convex, it is not easy to optimize it. For instance, when $f=f_e$,
usually (\ref{OPPA4}) is solved via the following semi-definite optimization problem with the introduction of slack symmetric $N\times N$ matrix variable
$\mathbf{T}$:
\[
\begin{array}{ll}
&\min_{\bx,\bT} \ \Tr(\bT)+\mu(1/g^{(\kappa)}(\bx)-1/S)\\
\mbox{s.t.}&\bx\in\polyD, (\ref{obser})/(\ref{depthone}), (\ref{quad2b}),
\begin{bmatrix}
    \clR^{-1}_e(\bx) & I_N \\
   I_N& \bT   \end{bmatrix}\succeq 0,
\end{array}
\]
which is not scalable to $\bx$.  For $f=-f_{MI}$, (\ref{OPPA4}) is
\[
\begin{array}{r}
\max_{\bx\in [0,1]^N} \ln|\clR_e^{-1}(\bx)|
-\mu(1/g^{(\kappa)}(\bx)-1/S)\\
\mbox{s.t.}\quad \bx\in\polyD, (\ref{quad2b}),
\end{array}
\]
with no known convex solver of polynomial complexity.

In the following, we propose a different approach to provide scalable iterations for (\ref{OPP1}).
Obviously, there is $\epsilon>0$ such that
\[
\clA_{\epsilon}:=B^T\Sigma_P^{-1} B -\epsilon\sum_{k\in\clN}H_k^TR_{w_k}^{-1}H_k\succ 0.
\]
For $f=f_e$, applying inequality (\ref{in2}) in the Appendix for
\begin{equation}\label{change}
A_0\rightarrow\clA_{\epsilon},
x_k\rightarrow x_k+\epsilon, \bar{x}_k\rightarrow x_k^{(\kappa)}+\epsilon,
\end{equation}
yields
$
f_{e}(\bx)\geq f_{e}^{(\kappa)}(\bx):=a_0^{(\kappa)}
+\ds\sum_{k\in\clN}\frac{a_k^{(\kappa)}}{x_k+\epsilon}$
for $0<a_0^{(\kappa)}:=\Tr((\clR_e(\bx^{(\kappa)}))^2\clA_{\epsilon})$ and
\[
\begin{array}{r}
0<a_k^{(\kappa)}:=(x^{(\kappa)}_k+\epsilon)^2\Tr((\clR_e(\bx^{(\kappa)}))^2H_k^TR_{w_k}^{-1}H_k ),\\
k\in\clN.
\end{array}
\]
Accordingly, initialized by a feasible point $\bx^{(0)}$ for (\ref{OPPA3}),
at the $\kappa$-th iteration for $\kappa=0, 1,\dots$,  we solve the following convex optimization problem to generate the
next iterative point $\bx^{(\kappa+1)}$, instead of (\ref{OPPA4}):
\begin{eqnarray}\label{OPPA4k}
\ds\min_{\bx} F_{\mu}^{(\kappa)}(\bx):=f_e^{(\kappa)}(\bx)+\mu(1/g^{(\kappa)}(\bx)1-1/S)\nonumber \\
\mbox{s.t.} \quad \bx\in\polyD, (\ref{obser})/(\ref{depthone}), (\ref{quad2b}).
\end{eqnarray}
Note that $F_{\mu}(\bx)\leq F_{\mu}^{(\kappa)}(\bx)$ $\forall\ \bx$, and
$F_{\mu}(\bx^{(\kappa)})=F^{(\kappa)}_{\mu}(\bx^{(\kappa)})$, and
$F^{(\kappa)}_{\mu}(\bx^{(\kappa+1)})<F^{(\kappa)}_{\mu}(\bx^{(\kappa)})$
(because $\bx^{(\kappa+1)}$ and $\bx^{(\kappa)}$ are the optimal solution and a feasible point
for (\ref{OPPA4k})). Therefore,
\[
F_{\mu}(\bx^{(\kappa+1)})\leq F^{(\kappa)}_{\mu}(\bx^{(\kappa+1)})<F^{(\kappa)}_{\mu}(\bx^{(\kappa)})
=F_{\mu}(\bx^{(\kappa)}),
\]
i.e. $\bx^{(\kappa+1)}$ is a better
feasible point than $\bx^{(\kappa)}$ for (\ref{OPPA3}). For a sufficient large $\mu>0$, $\tilde{g}(\bx^{(\kappa)})\rightarrow
0$ as well, yielding an optimal solution of the binary nonlinear  optimization problem (\ref{OPP1}) for the case $f=f_e$.
Algorithm \ref{alg1} provides a pseudo-code for the proposed computational procedure.\\
\begin{algorithm}
	\caption{Scalable Penalized MMSE  Algorithm } \label{alg1}
	\begin{algorithmic}[1]
		\State \textbf{Initialization.} {\sf Set $\kappa=0$. Take any feasible point $\bx^{(0)}\in (0,1)^N$ for
(\ref{OPPA3}). {\color{black}Choose $\mu$ such that $f_e(\bx^{(0)})$ and $(1/S-1/g(\bx^{(0)}))$ achieve similar magnitude.}}
		\State \textbf{Repeat}
		\State {\sf Solve the convex optimization problem (\ref{OPPA4k}) to generate the next feasible point
$\bx^{(\kappa+1)}$}.
		\State {\sf Set $\kappa := \kappa + 1$}.
		\State \textbf{Until} {\sf convergence}.
	\end{algorithmic}
\end{algorithm}

Analogously, based on  inequality (\ref{in4}) in the Appendix,  for $A_0$,
$x_k$, and $\bar{x}_k$ from (\ref{change}),
at the $\kappa$-th iteration we solve the following convex optimization problem to generate the
next iterative point $\bx^{(\kappa+1)}$, instead of (\ref{OPPA4}), when $f=-f_{MI}$:
\begin{eqnarray}\label{OPPA5k}
\ds\max_{\bx} \left[\alpha_0^{(\kappa)}-
\ds\sum_{k\in\clN}\frac{\alpha_k^{(\kappa)}}{x_k+\epsilon}-\mu(\frac{1}{g^{(\kappa)}(\bx)}-\frac{1}{S})\right] \nonumber \\
\mbox{s.t.} \quad \bx\in\polyD, (\ref{obser})/(\ref{depthone}), (\ref{quad2b}),\end{eqnarray}
for
\[
\begin{array}{lll}
\alpha_0^{(\kappa)}&:=& -\ln|\clR_e(\bx^{(\kappa)})|\\
&&+ \Tr(\clR_e(\bx^{(\kappa)})
(\sum_{k\in\clN}(\epsilon+x_k^{(\kappa)})H_k^TR_{w_k}^{-1}H_k)),\\
\alpha_k^{(\kappa)}&:=&(x^{(\kappa)}_k+\epsilon)^2\Tr(\clR_e(\bx^{(\kappa)})H_k^TR_{w_k}^{-1}H_k ), \\
&&k\in\clN.
\end{array}
\]
Algorithm \ref{alg2} is  a pseudo-code for solution of the binary nonlinear optimization problem (\ref{OPP1}) for the case $f=-f_{MI}$.\\

\begin{algorithm}
	\caption{Scalable Penalized MI Algorithm } \label{alg2}
	\begin{algorithmic}[1]
		\State \textbf{Initialization.} {\sf Set $\kappa=0$. Take any feasible point $\bx^{(0)}\in (0,1)^N$ for
(\ref{OPPA3}). {\color{black}Choose $\mu$ such that $f_{MI}(\bx^{(0)})$ and $(1/S-1/g(\bx^{(0)}))$ achieve similar magnitude.}}
		\State \textbf{Repeat}
		\State {\sf Solve the convex optimization problem (\ref{OPPA5k}) to generate the next feasible point
$\bx^{(\kappa+1)}$}.
		\State {\sf Set $\kappa := \kappa + 1$}.
		\State \textbf{Until} {\sf convergence}.
	\end{algorithmic}
\end{algorithm}
\section{Tailored path-following discrete optimization algorithms}
In this section, we address problem (\ref{OPP1})
without the observability constraint (\ref{obser})/(\ref{depthone}):
\begin{equation}\label{nOPP}
\ds\min_{\bx} f(\bx)\quad
\mbox{s.t.}\quad  \bx\in\clD_S.
\end{equation}
which was considered in \cite{CA06,KGW12}  for $f=f_e$  with the help of semi-definite relaxation (SDR). The reader is referred to \cite{Tuetal11} for capacity of SDR to address discrete optimization problems such as (\ref{nOPP}).
We now develop a simple but very efficient path-following discrete optimization algorithm that explores a simple structure
of the discrete constraint $\bx\in\clD_S$ to address (\ref{nOPP}).\\

\begin{mylem}\label{molem2} {\it $\clD_S$ is the set of vertices of $\polyD$.}\\
 \end{mylem}
\Prf
 For $\bx\in\clD_S$ define
\begin{equation}\label{accord1}
J(\bx)=\{k_1<k_2<....<k_S| x_{k_j}=1,\ j=1,2,...,S\}.
\end{equation}
Suppose $\bar{\bx}\in\clD_S$. It suffices to show that if $\bar{\bx}= \mu \ba+ (1-\mu) \bb$ for $\ba, \bb \in \polyD$
 and $0<\mu <1$ then $\ba=\bb=\bar{\bx}$. Indeed,
 for $i\in J(\bar{\bx})$ we have $\bar{x}_i=1= \mu a_i+ (1-\mu)b_i$ and since $a_i\in [0,1]$ and $b_i\in [0,1]$
it follows that $a_i=b_i=1$. For $i\notin J(\bar{\bx})$ we have $\bar{x}_i=0=\mu a_i+ (1-\mu)b_i$ and since $a_i\in [0,1]$,
and $b_i\in [0,1]$ it follows that $a_i=b_i=0.$ Hence $\ba=\bb=\bar{\bx}$ as asserted. \qed \\
Recall that point $\bx$ is a vertex neighbouring the vertex $\bar{\bx}$ if and only if there exists a pair $i$ and $j$ with $i\in \{S+1,\ldots,N\}$  and $j\in \{1,\ldots,S\}$ such that
$x_i= 1, x_j=0$ and $x_k= \bar{x}_k=1$ for all $k\in\clN\setminus\{j\}$ and $x_k= \bar{x}_k=0$ for all $k\in \{N+1,\ldots,M\}\setminus \{i\}.$

A $\bar{\bx}\in\clD_S$ is a minimizer of $f$ over $\polyD$ if and only if $f(\bar{\bx})\le f(\mathbf{v})$ for every $\mathbf{v}\in\clD_S$ neighbouring $\bar{\bx}$.

\begin{algorithm}
\caption{Path-following discrete optimization algorithm} \label{pfalg}
\begin{algorithmic}
\State{\it Initialization.} {\sf Start from a $\bx^{(0)}\in\clD_S$. Set $\kappa=0$}.
\State{\it $\kappa$-th iteration.} {\sf If there is a $\bar{\bx}\in\clD_S$ neighbouring $\bx^{(\kappa)}$ such that
$f(\bar{\bx})<f(\bx^{(\kappa)})$ then reset $\kappa+1\rightarrow \kappa$ and $\bx^{(\kappa)}\rightarrow
\bar{\bx}$.  Otherwise, if $f(\bx)\geq f(\bx^{(\kappa)})$
for all $\bx\in\clD_S$ neighbouring $\bx^{(\kappa)}$ then stop: $\bx^{(\kappa)}$ is a local optimal solution of (\ref{OPP1}).}
\end{algorithmic}
\end{algorithm}
 The proposed Algorithm \ref{pfalg} looks like the Dantzig simplex method for linear programming, which is of the 20th century's top ten algorithms \cite{C00} although
its polynomial complexity cannot be proved (in contrast to the polynomial complexity of
the interior points methods for linear programming).\footnote{Conceptually, Dantzig simplex method is very simple:
starting from any vertex of a simplex it moves to a better neighbouring vertex until there
is no better neighbouring vertex found}
Based on this powerful algorithm, we propose Algorithm \ref{ifalg} for
the following problem of choosing the minimum number of PMUs to satisfy MMSE or MI constraint:
\begin{equation}\label{minp}
\min_{\bx}\ \sum_{k\in\clN}x_k\ :\  \bx\in \{0,1\}^N, f(\bx)\leq \epsilon.
\end{equation}
\begin{algorithm}
\caption{Iterative Procedure} \label{ifalg}
\begin{algorithmic}
\State{\it Initialization.} {\sf Start from $1<S_0<N$ and use Algorithm \ref{pfalg} to find the optimal solution $\bx^{(0)}$ of
(\ref{nOPP}) for $S=S_0$.}
\State{\it $\kappa$-th iteration.} {\sf Reset $S\rightarrow S-1$ if $f(\bx_{opt})<\epsilon$ and
$S\rightarrow S+1$ if $f(\bx_{opt})>\epsilon$}.
\State {\sf Set $\kappa := \kappa + 1$}.
		\State \textbf{Until} $f(\bx^{(\kappa)})\leq \epsilon$ {\sf but} $f(\bx^{(\kappa-1)})>\epsilon$.
\end{algorithmic}
\end{algorithm}
\vspace*{-0.5cm}
\section{Simulation results}
In the simulation, the real power injections $P$ are
normally distributed and independent across different buses \cite{AWJ05}.
Similarly to the simulation setup in \cite{Lietal13}, the mean vector of real power injection
$u_p=(u_p(1), \dots, u_p(N))^T$ is obtained by properly scaling the power profiles in \cite{ZMT11}, while the diagonal  entries of power injection covariance matrix are assumed to be 10\% of the mean values, i.e. $\Sigma_P$ is
 a diagonal matrix with diagonal entries $\Sigma_P(k,k)= 0.1 u_p(k)$. The deviation of measurement noise for bus voltage and current branch are set as $r_k= 0.01$ and $\rho_k= 0.02$, respectively. All algorithms are solved by Matlab on a Core i7-7600U processor. Sedumi\cite{S98} interfaced by CVX is used to solve the convex optimization problems (\ref{OPPA4k}) and (\ref{OPPA5k}). The commonly used benchmark power networks IEEE 30-bus, IEEE 39-bus, IEEE 57-bus and IEEE 118-bus
with their structure and susceptance matrix obtained from Matpower \cite{ZMT11} are tested.

It is observed in \cite{G208} that the minimum number of PMUs for the network complete observability (CO) or depth-of-one unobservability (DoOU) can be found by solving the following binary linear program
\begin{equation}\label{inp1}
\min_{\bx}\ \sum_{k\in\clN}x_k\ :\  \bx\in \{0,1\}^N, (\ref{obser})/(\ref{depthone}).
\end{equation}
Table \ref{Case_inform} provides the minimum number of PMUs needed for the network's  CO and DoOU (obtained by solving (\ref{inp1} by
CPLEX  \cite{CPLEX})
given in the third and fourth columns.
\begin{table}[h]
    \centering
    \caption{The minimum number of PMUs needed for two observability conditions}
    \begin{tabular}{cccc}
    \hline \hline
    IEEE & \# Branch& \# PMUs for CO& \# PMUs for DoOU\\
    \hline
    30-bus & 41 & 10 & 4 \\
    39-bus & 46 &  13 & 7 \\
    57-bus & 80 &  17 & 11 \\
    118-bus & 186 & 32 & 18 \\
    \hline \hline
    \end{tabular}
\label{Case_inform}
\end{table}

\begin{figure}[h]
\centering
\includegraphics[width=1.0 \columnwidth]{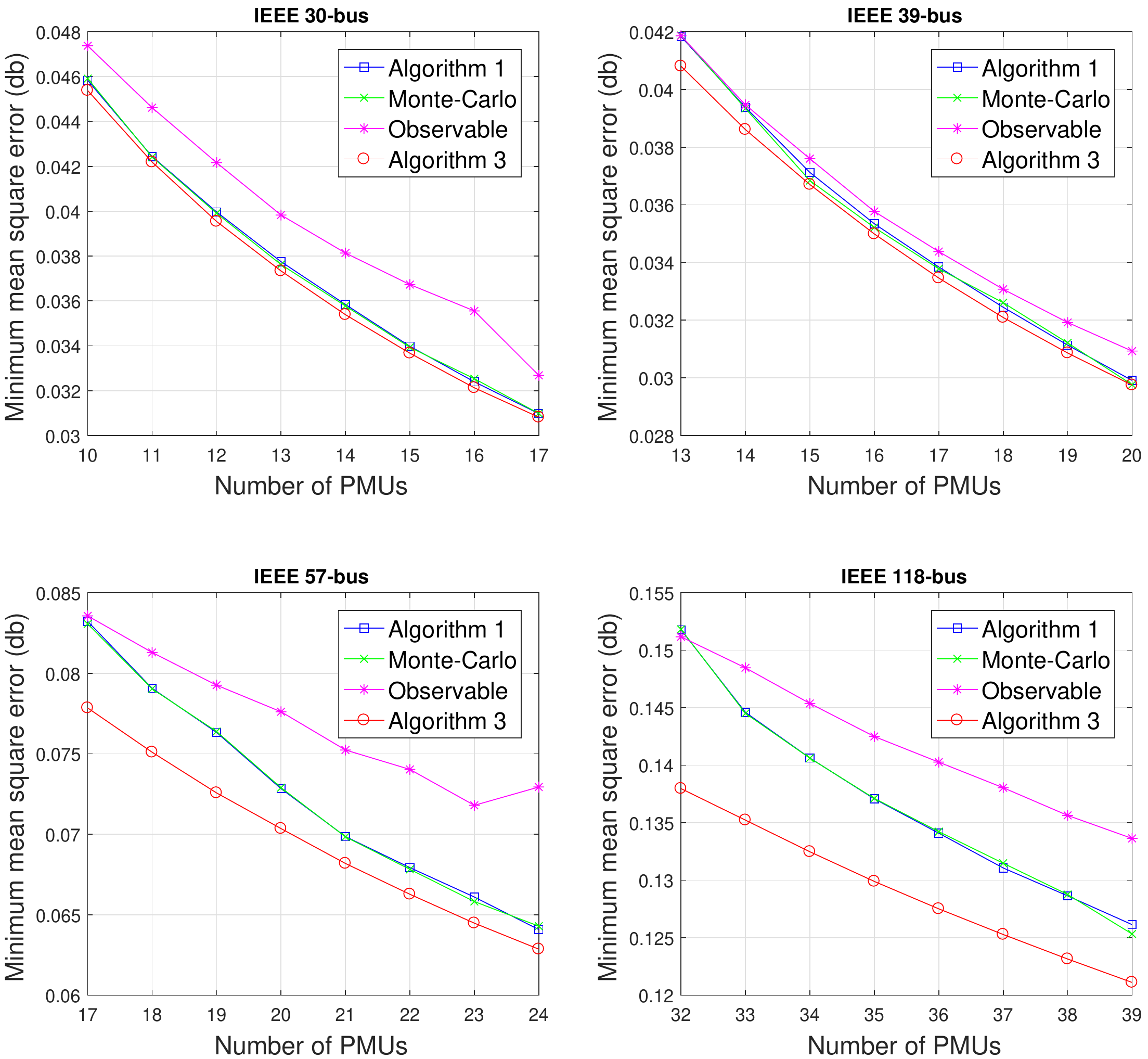}
\caption{MMSE by different methods}
\label{MMSE_30_39_57_118}
\end{figure}

\begin{figure}[h]
\centering
\includegraphics[width=1.0 \columnwidth]{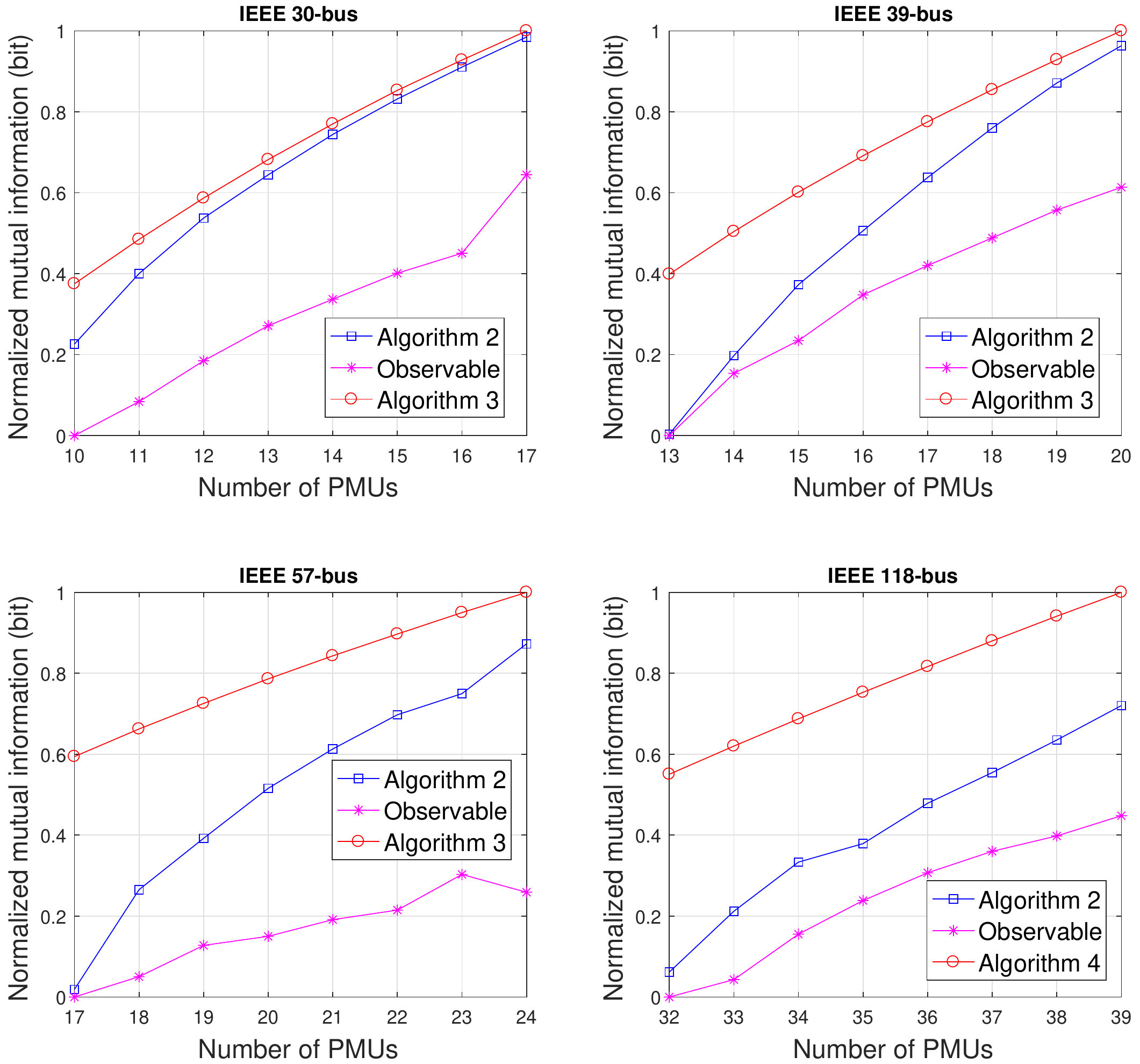}
\caption{MI by different methods}
\label{MI_30_39_57_118}
\end{figure}

Fig.\ref{MMSE_30_39_57_118} depicts the MMSE obtained by different methods versus the number of placed PMUs.
The curve  "Algorithm \ref{alg1}" is the theoretical MMSE by solving problem (\ref{OPPA4k})
under the constraint (\ref{obser}) of the complete observability, while the curve "Monte-Carlo" is obtained through  Monte-Carlo simulation. The MMSEs by Algorithm \ref{alg1} and Monte-Carlo simulation are seen consistent with the
increase in the number of placed PMUs leading to a better MMSE.
The curve "Observable" is the MMSE at feasible points for (\ref{OPPA4k}) that is found by CPLEX \cite{CPLEX}.
Algorithm \ref{alg1} is seen to achieve much better MMSE. The last curve "Algorithm 3" is the MMSE
by solving (\ref{nOPP}) by Algorithm \ref{pfalg}. Obviously, the Algorithm \ref{pfalg} achieves
better MMSE due to the absence of constraints (\ref{obser}) and (\ref{depthone}).
The curves in Fig. \ref{MI_30_39_57_118} provide normalized MI results in a similar format to
Fig.\ref{MMSE_30_39_57_118}. The capability and efficiency of Algorithm \ref{alg2} and Algorithm \ref{pfalg} to obtain informative PMU placements are quite clear.

Table \ref{Alg1_2_3} provides numerical details of Algorithm \ref{alg1}, Algorithm \ref{alg2} and Algorithm \ref{pfalg}.
The value of the penalized parameter $\mu$ in implementing Algorithm \ref{alg1} and Algorithm \ref{alg2} is given by the second and fourth column, while  the average CPU time
is given by the third and fifth column. The last two columns provide average CPU time
by Algorithm \ref{pfalg} in solving MMSE and MI.
Algorithm \ref{pfalg} needs much less time for small-scale networks but
its computational cost increases dramatically with the growth of network size.
On the other hand, the CPU time of Algorithm \ref{alg1} and Algorithm \ref{alg2} increases moderately
when the size of networks grows, demonstrating their scalability and
superiority in addressing  large-scale networks.
\begin{table}[h]
    \centering
    \caption{Numerical details of Algorithm \ref{alg1}, Algorithm \ref{alg2} and Algorithm \ref{pfalg}}
    \begin{tabular}{ccccccc}
    \hline \hline
    \multirow{2}{*}{IEEE } & \multicolumn{2}{c}{Alg. \ref{alg1}} & \multicolumn{2}{c}{Alg. \ref{alg2}} & \multicolumn{2}{c}{CPU (s) of Alg. \ref{pfalg}} \\ \cline{2-7}
    & $\mu$ & Avg. T. (s) & $\mu$ & CPU (s) & MMSE & MI \\
    \hline
    30-bus& 0.1 & 65.78 & 1 & 62.94 & 4.01 & 3.17\\
    39-bus& 0.1 & 79.73 & 1 & 77.25 & 11.58 & 7.98\\
    57-bus& 1 & 80.47 & 10 & 81.14 & 49.09 & 46.03\\
    118-bus& 1 & 216.31 & 10 & 193.24& 1222.11 & 2142.08\\
    \hline \hline
    \end{tabular}
\label{Alg1_2_3}
\end{table}

\begin{figure}[h]
\centering
\includegraphics[width=1.0 \columnwidth]{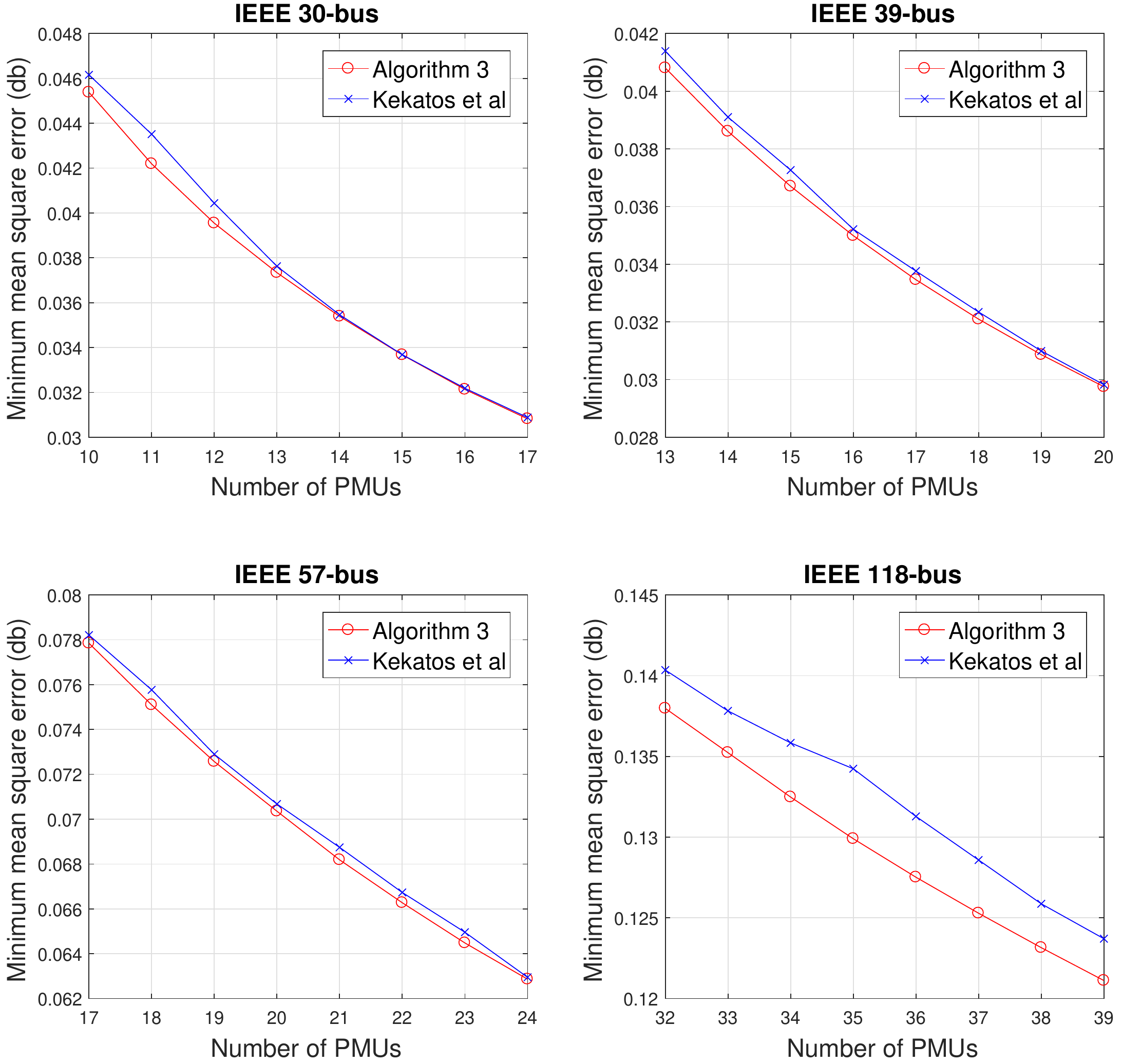}
\caption{MMSE found by Algorithm 3 and by \cite{KGW12}}
\label{Alg3_Kekatoes}
\end{figure}

For problem (\ref{nOPP}), Kekatos et al \cite{KGW12} relaxed the integer constraint $\bx\in \{0,1\}^N$ to
the box constraint $\bx\in [0,1]^N$ to formulate a convex problem and then round the $S$ largest values of the solution
of this convex program to $1$. Obviously, their solution is hardly optimal in any sense.
Fig. \ref{Alg3_Kekatoes} compares the  MMSE values of problem (\ref{nOPP}) founded by  Algorithm \ref{pfalg} and
by Kekatos et al \cite{KGW12}. The former clearly outperforms the latter, especially for large scale networks.

\begin{figure}[h]
\centering
\includegraphics[width=1.0 \columnwidth]{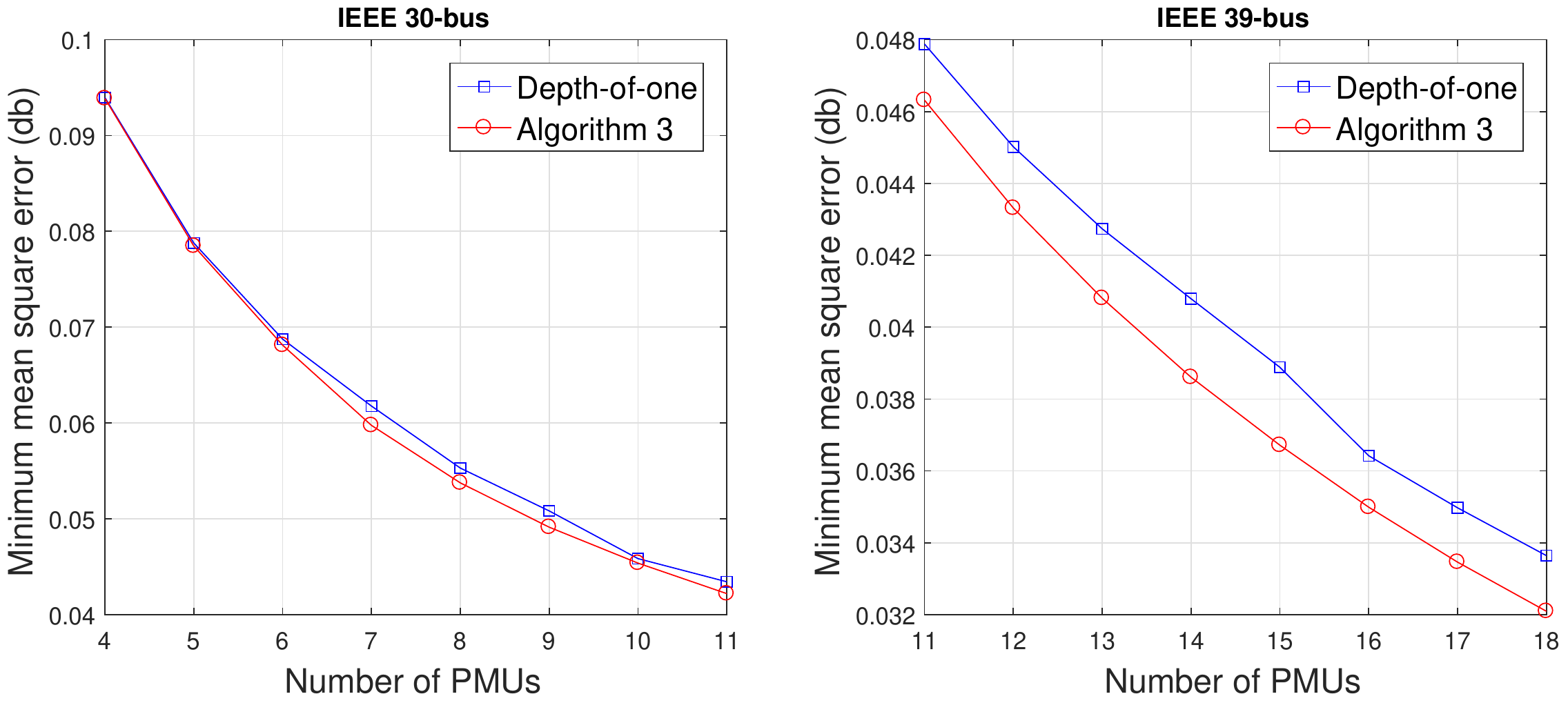}
\caption{MMSE found by Algorithm \ref{alg1} under depth-of-one unobservability condition and Algorithm \ref{pfalg} without any observability constraints}
\label{Depth_of_one_case30_39}
\end{figure}

\begin{figure}[h]
\centering
\includegraphics[width=0.9 \columnwidth]{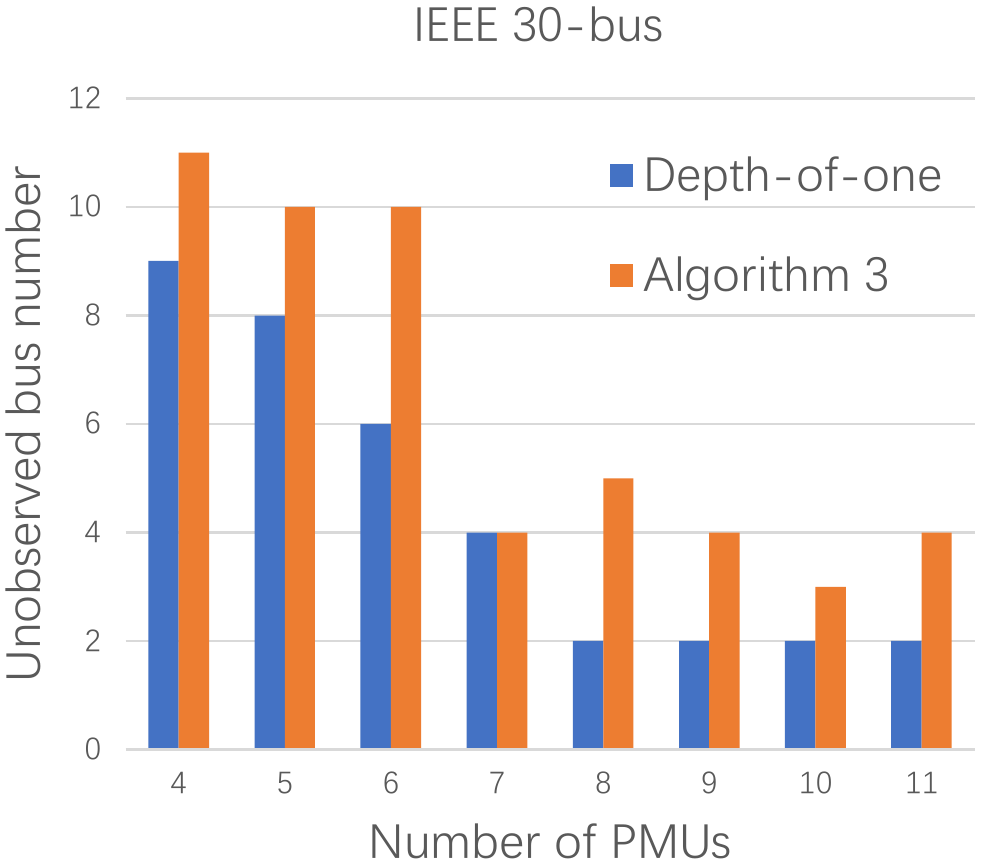}
\caption{Number of buses left unobserved by Algorithm \ref{alg1} under depth-of-one unobservability condition and Algorithm \ref{pfalg} without any observability constraints for IEEE 30-bus network}
\label{IEEE_30}
\end{figure}

Due to space limitation, only IEEE 30-bus and IEEE-39 networks are selected for
MMSE results solved by Algorithm \ref{alg1} under the constraint (\ref{depthone})
of depth-of-one unobservability. Fig. \ref{Depth_of_one_case30_39} provides  MMSE performance obtained via
Algorithm \ref{alg1} (under the constraint (\ref{depthone}))  and Algorithm \ref{pfalg} (without any observability constraints), while Fig. \ref{IEEE_30} provides the number of bus left unobservable (for IEEE 30-bus).
As expected, Algorithm \ref{pfalg} achieves a better MMSE but leaves more buses unobservable because it sacrifices
buses to  achieve the averaged performance.

\begin{figure}[h]
\centering
\includegraphics[width=1.0 \columnwidth]{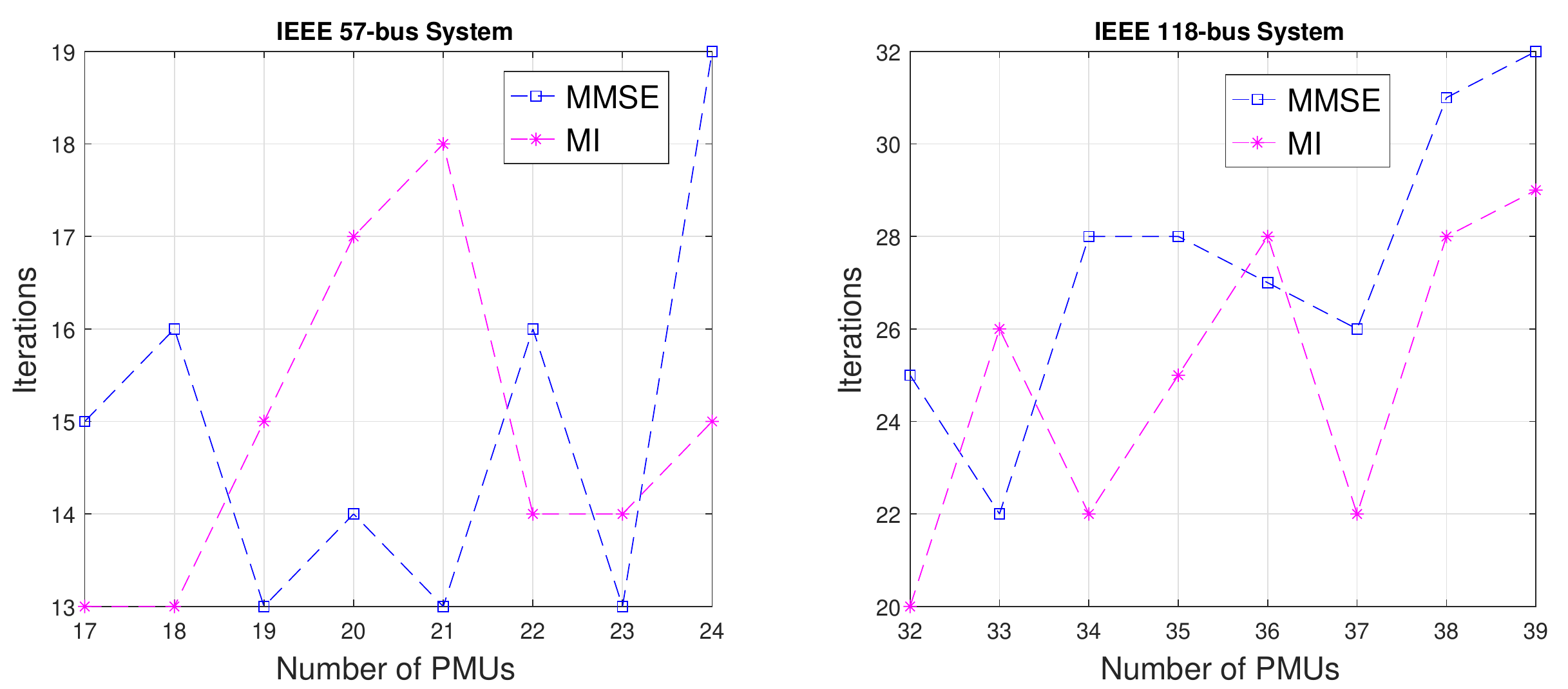}
\caption{Number of iterations required for the convergence of Algorithm 3}
\label{GOA_Ite_57_118}
\end{figure}

For IEEE 57-bus network and IEEE 118-bus network, Fig.\ref{GOA_Ite_57_118} presents the number of iterations
needed for the convergence of Algorithm 3 for MMSE and MI, respectively.

\begin{figure}[h]
\centering
\includegraphics[width=1.0 \columnwidth]{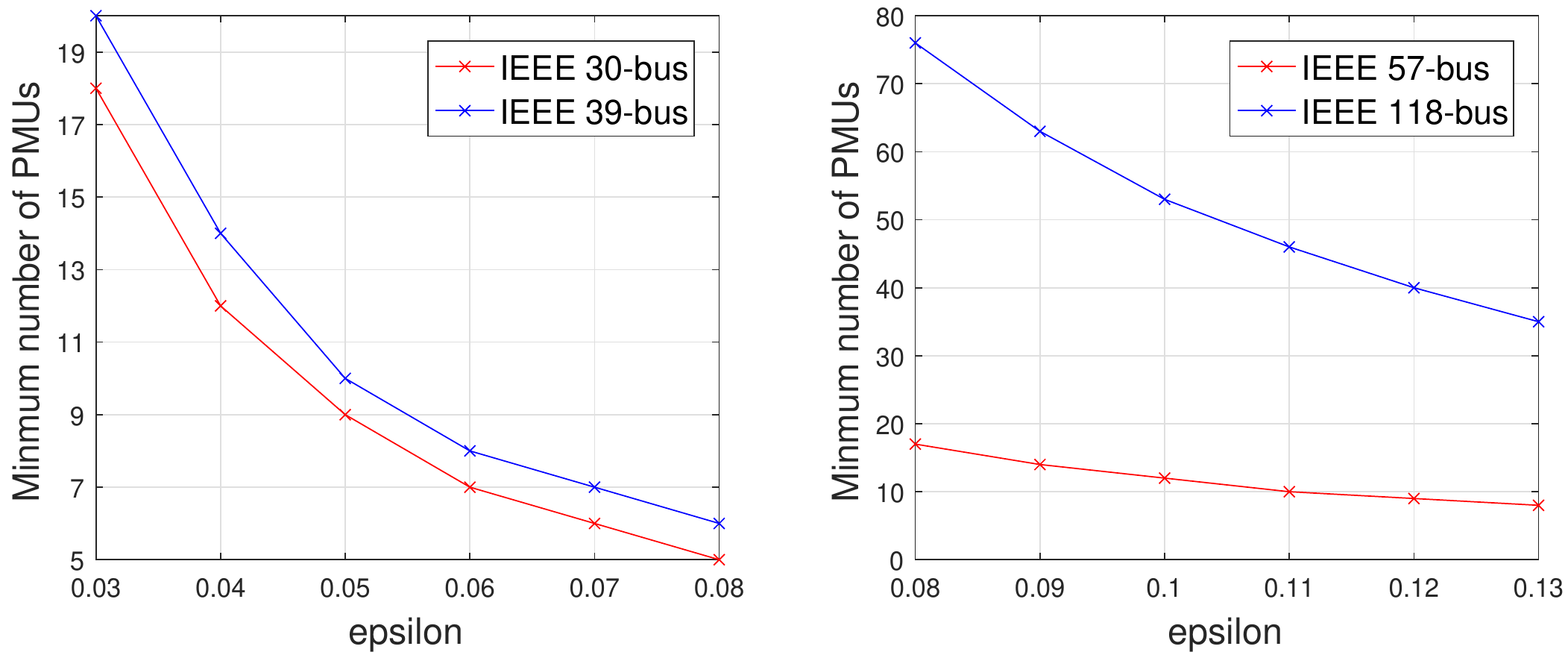}
\caption{Minimum number of PMUs required versus different values of tolerance level $\epsilon$ for MMSE}
\label{Alg4_MMSE}
\end{figure}

Given different tolerances $\epsilon$, the required minimum number of PMUs can be obtained by Algorithm \ref{ifalg}. For the
case of $f=F_e$, the results are presented in Fig.\ref{Alg4_MMSE}.
\section{Conclusions}
In this paper, we have considered PMU placement optimization to minimize the mean squared error or maximize
the mutual information between the measurement outputs and phasor states under a fixed number of PMUs and different observability conditions. These binary optimization problems are very computationally challenging due to high nonlinearity of the objective functions. Nevertheless, we have developed the scalable algorithms for their computational solution, which
result at least in local optimal solutions. We also developed extremely efficient algorithms of very low computational
complexity for cases of absent observability. The viability of our proposed algorithms has been confirmed through
simulations with benchmark IEEE grids. The algorithmic developments for PMU  placement optimization involving other practical
constraints such as branch outages are under way.
\section*{Appendix: Fundamental Inequalities}
Let $\mathbb{R}^N_+:=\{ x\in\mathbb{R}^N : x_k\geq 0, k\in\clN\}$ and $\inte(\mathbb{R}^N_+):=\{ x\in\mathbb{R}^N : x_k> 0, k\in\clN\}$. For $A_0\succ 0$ and $A_k\succeq 0$, $k\in\clN$ let
$
\Phi(\bx):=(A_0+\ds\sum_{k\in\clN}\frac{1}{x_k}A_k)^{-1},
$
and
$
\Psi(\bx):=(A_0+\ds\sum_{k=1}^Nx_kA_k)^{-1}
$.
Recall the following result \cite[Th.1]{Beetal18}:\\

\begin{myth}\label{bin1}{\it
Function $\varphi(\bx)=\Tr(\Phi(\bx))$ is concave
in the domain $\inte(\mathbb{R}^N_+)$, so for all $\bx\in \inte(\mathbb{R}^N_+)$ and $\bar{\bx}\in \inte(\mathbb{R}^N_+)$
one has
\[
\begin{array}{lll}
\varphi(\bx)&\leq&\varphi(\bar{\bx})+\lan \nabla\varphi(\bar{\bx}), \bx-\bar{\bx}\ran\\
&=&\Tr\left(\Phi^2(\bar{\bx})A_0\right)+\ds
\sum_{k\in\clN}\frac{x_k}{\bar{x}_k^2}\Tr\left(\Phi^2(\bar{\bx})A_k \right).
\end{array}
\]
Therefore,
\begin{eqnarray}\label{in2}
\Tr(\Psi(\bx))&\leq&
\Tr\left((\Psi(\bar{\bx}))^2A_0\right)\nonumber\\
&&+\ds\sum_{k\in\clN}\frac{\bar{x}_k^2}{x_k}\Tr\left((\Psi(\bar{\bx}))^2A_k \right).
\end{eqnarray}
}
\end{myth}
Next,\\

\begin{myth}\label{cth1}{\it
For $A\succ 0$ function $\ln|A+H\bX^{-1}H^H|$ is convex in $\bX\succ 0$.}
\end{myth}

\Prf Since $
(A+H\bX^{-1}H^H)^{-1}=
A^{-1}-A^{-1}(H^HA^{-1}H+\bX)^{-1}A^{-1}
$,
by \cite[Appendix B]{Raetal14}, function
\[
f(\bX):=A^{-1}-A^{-1}(H^HA^{-1}H+\bX)^{-1}A^{-1}
\]
is concave, i.e.
$
f(\alpha\bX+\beta\bY)\succ \alpha f(\bX)+\beta f(\bY)\quad\ \forall \ \bX\succ 0, \bY\succ 0$,
and $\alpha\geq 0, \beta\geq 0, \alpha+\beta=1$. Therefore
$
\ln|f(\alpha\bX+\beta\bY)| \geq  \ln|\alpha f(\bX)+\beta f(\bY)|\geq \alpha\ln|f(\bX)|+\beta\ln|f(\bY)|$,
showing that $\ln|A+H\bX^{-1}H^H|^{-1}=-\ln|A+H\bX^{-1}H^H|$ is concave in $\bX$.\qed\\

The following Theorem is a direct consequence of Theorem \ref{cth1}.

\begin{myth}\label{bin1}{\it
Function $\phi(\bx)=-\ln|\Phi(\bx)|$ is convex
in the domain $\inte(\mathbb{R}^N_+)$, so for all $\bx\in \inte(\mathbb{R}^N_+)$ and $\bar{\bx}\in \inte(\mathbb{R}^N_+)$
one has
\[
\begin{array}{lll}
\phi(\bx)&\geq&\phi(\bar{\bx})+\lan \nabla\phi(\bar{\bx}), \bx-\bar{\bx}\ran\\
&=&-\ln|\Phi(\bar{\bx})|+\ds
\Tr\left((\Phi(\bar{\bx}))^{-1}(\sum_{k\in\clN}\frac{1}{\bar{x}_k}A_k)\right)\\
&&\ds-\sum_{k\in\clN}\frac{x_k}{\bar{x}_k^2}\Tr\left((\Phi(\bar{\bx}))^{-1}A_k \right).
\end{array}
\]
Therefore,
\begin{eqnarray}\label{in4}
-\ln|\Psi(\bx)|\geq -\ln|\Psi(\bar{\bx})| +\Tr\left(\Psi(\bar{\bx})(\sum_{k\in\clN}\bar{x}_kA_k)\right) \nonumber\\
-\ds\sum_{k\in\clN}\frac{\bar{x}_k^2}{x_k}\Tr\left(\Psi(\bar{\bx})A_k \right).
\end{eqnarray}
}
\end{myth}
\bibliographystyle{ieeetr}
\bibliography{OPP_BIB}
\end{document}